\documentclass[preprint2]{aastex}
\usepackage{amssymb,epsfig,wrapfig,graphics,aas_macros}
\usepackage[dvips]{color}
\usepackage[footnotesize]{caption}
\def\comment#1{{}}
\newlength{\cvindent}
\newlength{\cvhang}

\newlength{\refindent}
\begin{document}
\title{Section on Extragalactic Science Topics of the White Paper on the Status and Future of Ground-Based TeV Gamma-Ray Astronomy}
\author{\large
H.~Krawczynski (Wash.\ University in St. Louis, Physics Department and McDonnell Center for the Space Sciences),
A.~M.~Atoyan (Montreal University),
M.~Beilicke (Wash.\ University in St. Louis, Physics Department and McDonnell Center for the Space Sciences),
R.~Blandford (Kavli Institute for Particle Astrophysics and Cosmology, Stanford Linear Accelerator Center),
M.~Boettcher (Ohio University),
J.~Buckley (Wash.\ University in St. Louis, Physics Department and McDonnell Center for the Space Sciences),
A.~Carrami\~nana (Instituto Nacional de Astrof\'{\i}sica, \'{O}ptica y Electr\'{o}nica, M\'{e}xico),
P.~Coppi (Yale University),
C.~Dermer (U.S. Naval Research Laboratory),
B.~Dingus (Los Alamos),
E.~Dwek (NASA Goddard Space Flight Center),
A.~Falcone (Pennsylvania State University),
S.~Fegan (University of California, Los Angeles, now at: Institut National de la Physique Nucleaire),
J.~Finley (Purdue University),
S.~Funk (Kavli Institute for Particle Astrophysics and Cosmology, 
Stanford Linear Accelerator Center),
M.~Georganopoulos (University of Maryland, NASA Goddard Space Flight Center),
J.~Holder (University of Delaware),
D.~Horan (Argonne National Laboratory, now at: Institut National de la Physique Nucleaire),
T.~Jones (University of Minnesota),
I.~Jung (Wash.\ University in St. Louis, Physics Department and McDonnell Center for the Space Sciences, now at: Universit\"at 
Erlangen-N\"urnberg),
P.~Kaaret (The University of Iowa),
J.~Katz (Wash.\ University in St. Louis, Physics Department and McDonnell Center for the Space Sciences),
F.~Krennrich (Iowa State University),
S.~LeBohec (University of Utah),
J.~McEnery (NASA Goddard Space Flight Center),
R.~Mukherjee (Columbia University),
R.~Ong (University of California, Los Angeles),
E.~Perlman (Flordia Institute of Technology),
M.~Pohl (Iowa State University),
S.~Ritz (NASA Goddard Space Flight Center),
J.~Ryan (University of New Hampshire),
G.~Sinnis (Los Alamos National Laboratory),
V.~Vassiliev (University of California, Los Angeles),
M.~Urry (Yale University),
T.~Weekes (Smithsonian Astrophysical Observatory).}
\begin{abstract}
This is a report on the findings of the extragalactic science working group for the white paper on the status and 
future of TeV gamma-ray astronomy. The white paper was commissioned by the American Physical Society, and the 
full white paper can be found on astro-ph (arXiv:0810.0444). This detailed section discusses extragalactic 
science topics including active galactic nuclei, cosmic ray acceleration in galaxies, galaxy clusters and 
large scale structure formation shocks, and the study of the extragalactic infrared and optical background 
radiation. The scientific potential of ground based gamma-ray observations of Gamma-Ray Bursts and dark 
matter annihilation radiation is covered in other sections of the white paper.
\vspace*{1.5cm}
\end{abstract}
\maketitle
\vspace*{1cm}

\section{Introduction}
A next-generation gamma-ray experiment will make extragalactic 
discoveries of profound importance. Topics to which gamma-ray observations can make unique
contributions are the following:
(i) the environment and growth of Supermassive Black Holes; (ii) the acceleration
of cosmic rays in other galaxies; (iii) the largest particle accelerators in
the Universe, including radio galaxies, galaxy clusters, 
and large scale structure formation shocks; (iv) study of the 
integrated electromagnetic luminosity of the Universe and
intergalactic magnetic field strengths through processes including
pair creation of TeV gamma rays interacting with infrared photons
from the Extragalactic Background Light (EBL).

The scientific potential of ground based observations of gamma-ray bursts 
and $\gamma$-rays from dark matter annihilation processes is described 
in other sections of the white paper \cite{wp}.
\section{Gamma-ray observations of supermassive black holes}
\begin{figure*}[t!h]
\includegraphics[angle=270,width=6.5in]{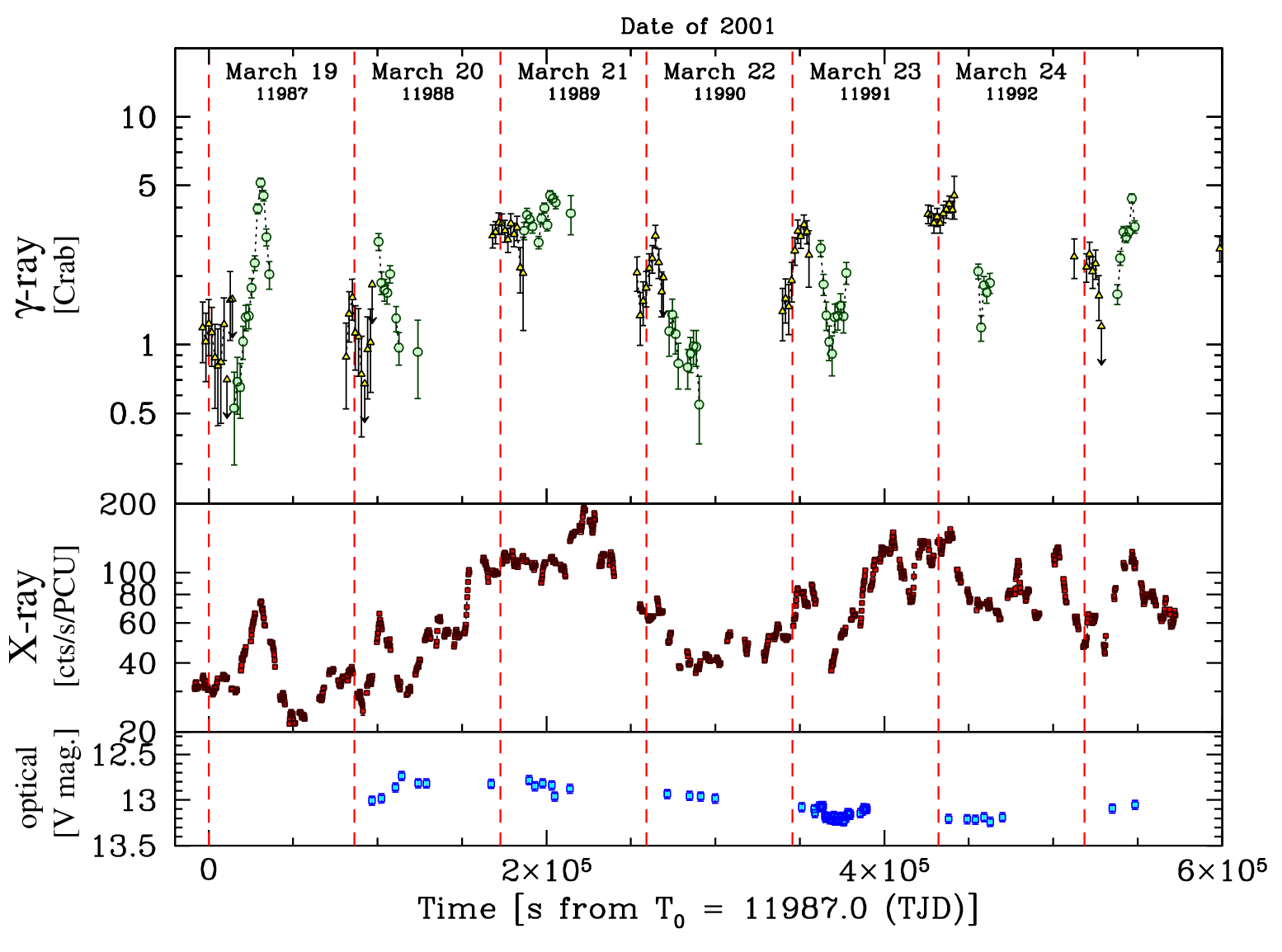} 
\caption{\label{agn} Results from 2001 Rossi X-ray Timing Explorer (RXTE) 2-4 keV X-ray and
Whipple (full symbols) and HEGRA (open symbols) gamma-ray observations 
of Mrk 421 in the year 2001 \cite{Foss:08}.
The X-ray/gamma-ray fluxes seem to be correlated. However, the
interpretation of the data is hampered by the sparse coverage at
TeV gamma rays.}
\end{figure*}

Supermassive black holes (SMBH) have masses between a million and several
billion solar masses and exist at the centers of galaxies. 
Some SMBHs, called Active Galactic Nuclei (AGN) are strong emitters
of electromagnetic radiation. 
Observations with the {\it EGRET Energetic Gamma-Ray Experiment Telescope}
on board of the Compton Gamma-Ray Observatory (CGRO) revealed that
a certain class of AGN known as blazars are powerful and variable 
emitters, not just at radio through optical wavelengths, but also 
at $\ge$100 MeV gamma-ray energies \cite{Hartman1999}. 
EGRET largely detected quasars, the most powerful blazars in the Universe.
Observations with ground-based Cherenkov telescopes showed that blazars 
emit even at TeV energies \cite{Punch1992}. In the meantime, more than
twenty blazars have now been identified as sources of $>$200~GeV gamma rays 
with redshifts ranging from 0.031 (Mrk 421) \cite{Punch1992} 
to 0.536 (3C~279) \cite{2007arXiv0709.1475T} \footnote{Up-to-date 
lists of TeV $\gamma$-ray sources can be found at the web-sites:
http$://$tevcat.uchicago.edu and 
http$://$www.mpp.mpg.de$/\sim$rwagner$/$sources$/$.}. 
Most TeV bright sources are BL Lac type objects, 
the low power counterparts of the quasars detected by EGRET. 
The MeV to TeV gamma-ray emission from blazars is commonly thought to 
originate from highly relativistic collimated outflows (jets) from mass 
accreting SMBHs that point at the observer \cite{Tavecchio2005,Krawczynski2006}.
The only gamma-ray emitting AGN detected to date that are not blazars are the 
radio galaxies Centaurus A \cite{Sreekumar1999} and M87 \cite{Aharonian2007M87}.
The observation of blazars in the gamma-ray band has had a major
impact on our understanding of these sources. The observation of rapid
flux variability on time scales of minutes together with 
high gamma-ray and optical fluxes
\cite{1996Natur.383..319G,2007ApJ...664L..71A} implies that the accreting 
black hole gives rise to an extremely relativistic jet-outflow with a bulk 
Lorentz factor exceeding 10, most likely even in the range between 10 and 50 
\cite{2001ApJ...559..187K,2008MNRAS.384L..19B}.
Gamma-ray observations thus enable us to study plasma which moves with
$\ge$99.98\% of the speed of light.
Simultaneous broadband multiwavelength observations of blazars have
revealed a pronounced correlation of the X-ray and TeV gamma-ray
fluxes \cite{1996ApJ...472L...9B,1996ApJ...470L..89T,2000A&A...353...97K,Foss:08}. 
The X-ray/TeV flux correlation (see Fig. \ref{agn}) suggests that the emitting particles
are electrons radiating synchrotron emission in the radio to X-ray
band and inverse Compton emission in the gamma-ray band. 

Blazars are expected to be the most copious extragalactic sources
detected by ground-based IACT arrays like VERITAS and by the satellite
borne gamma-ray telescope Fermi. For extremely strong
sources, IACT arrays will be able to track GeV/TeV fluxes on 
time scales of seconds and GeV/TeV energy spectra on time scales 
of a few minutes.
Resolving the spectral variability during individual strong flares
in the X-ray and gamma-ray bands should lead to the 
unambiguous identification of the emission mechanism. 
The present generation of IACTs will be able to track spectral
variations only for a very small number of sources and only during
extreme flares. The next-generation gamma-ray experiments will be able
to do such studies for a large number of sources on a routine basis.
Sampling the temporal variation of broadband energy spectra from a few
tens of GeV to several TeV will allow us to use blazars as precision
laboratories to study particle acceleration and turbulence in
astrophysical plasmas, and to determine the physical parameters
describing a range of different AGN.
The observations of blazars hold the promise to reveal details about the inner
workings of AGN jets. Obtaining realistic estimates of the power in
the jet, and the jet medium will furthermore constrain the origin of
the jet and the nature of the accretion flow.

Recently, spectracular results have been obtained by combining monitoring VLBA, 
X-ray and TeV $\gamma$-ray observations. This combination has the potential to pinpoint 
the origin of the high energy emission based on the high resolution radio images, 
and thus to directly confirm or to refute models of jet formation.
For example, radio VLBA, optical polarimetry, X-ray and TeV $\gamma$-ray observations 
of the source BL Lac seem to indicate that a plasma blob first detected with the 
VLBA subsequently produces an X-ray, an optical and a $\gamma$-ray flare \cite{2008Natur.452..966M}.
A swing of the optical polarization seems to bolster the case for a helical magnetic field
as predicted by magnetic models of jet formation and acceleration.
Presently such observations are extremely difficult as the current instruments 
can detect sources like M 87, BL Lac, W Com only in long observations or during extreme flares.
Next-generation $\gamma$-ray instruments will allow us to study the correlation 
of fast TeV flares and radio features on a routine basis.

In addition to ground-based radio to optical coverage, 
several new opportunities might open up within the next decade.
The Space Interferometry Mission (SIM) will be able to image emerging
plasma blobs with sub milli-arcsec angular resolution \cite{Unwi:02}. 
The center may be located with an accuracy of 
a few  micro-arcsec. For a nearby blazar at z=0.03, 1 milli-arcsec 
corresponds to a projected distance of 0.6 pc. The SIM observations 
could thus image the blobs that give rise to the flares detected in 
the gamma-ray regime. Joint X-ray/radio interferometry observations already give some tentative evidence for the emergence of radio blobs 
correlated with X-ray flares. 
If a Black Hole Finder Probe like the Energetic X-ray Imaging Space 
Telescope (EXIST) \cite{Grin:05} will be launched, it would provide reliable
all-sky, broad-bandwidth (0.5-600 keV), and high-sensitivity X-ray coverage 
for all blazars in the sky. EXIST's full-sky sensitivity would be 
2 $\times$ 10$^{-12}$ ergs cm$^{-2}$ s$^{-1}$ for 1 month of integration.
For bright sources, EXIST could measure not only flux variations but also
the polarization of hard X-rays. 
Opportunities arising from neutrino coverage will be described below.

At the time of writing this white paper, the Fermi gamma-ray telescope is in the process
of detecting a few thousand blazars. The source sample will make it possible to
study the redshift dependent luminosity function of blazars, although
the identification of sources with optical counterparts may be
difficult for the weaker sources of the sample, owing to Fermi's
limited angular resolution. Another important task for the next-generation 
instrument will be to improve on the Fermi localization accuracies, and thus 
to identify a large number of the weaker Fermi sources.

Independent constraints on the jet power, kinematics, and emission 
processes can be derived from GeV-TeV observations of the large scale 
(up to hundreds of kpc) jets recently detected by Chandra. Although such 
large scale jets will not be  spatially resolved, the fact that the 
gamma-ray emission from the quasar core is highly variable 
permits  us to set upper 
limits to the steady GeV-TeV large scale jet emission \cite{2006ApJ...653L...5G}.
In the case of the relatively nearby 3C 273, for example, the electrons that 
produce the large scale jet IR emission will also produce a flat GeV component. 
The fact that this emission is weaker than the EGRET upper limit constrains 
the Doppler factor of the large scale jets to less than 12, a value that can 
be pushed down to 5 with Fermi observations.  Such low values of delta have 
implications on the nature of the large scale jet X-ray emission observed 
by Chandra. In particular, they disfavor models in which the X-ray emission is 
inverse Compton scattering of the cosmic microwave background (CMB), because 
the jet power required increases beyond the so-called Eddington luminosity, 
thought by many to be the maximum luminosity that can be channeled continuously 
in a jet.  A synchrotron interpretation for the X-ray emission, requiring 
significantly less jet power, postulates a population of multi-TeV electrons 
that will unavoidably  up-scatter the CMB to TeV energies. The existing 3C~273 
shallow HESS upper limit constrains the synchrotron interpretation to Doppler 
factors less than 10. Combining deep TeV observations with a next-generation
experiment with Fermi observations holds the promise of  confirming 
or refuting the synchrotron interpretation and constraining the jet power. 

Whereas the X-ray/gamma-ray correlation favors leptonic models with electrons
as the emitters of the observed gamma-ray emission, hadronic models are not ruled out.
In the latter case, the high-energy component is synchrotron emission, either 
from extremely high-energy (EHE) protons \cite{Aharonian2000,Muecke2001,Muecke2003}, 
or from secondary $e^+/e^-$ resulting from synchrotron and pair-creation cascades 
initiated by EHE protons \cite{Mannheim1993} or high-energy electrons or photons 
\cite{Lovelace1979,Burns1982,Blandford1995,Levinson1995}. 
If blazars indeed accelerate UHE protons, it might even be possible to correlate their
TeV gamma-ray emission with their flux of high-energy neutrinos detected by
the IceCube detector \cite{Halz:05}. The high sensitivity of a next-generation
ground-based experiment would be ideally suited to perform such multi-messenger
studies.

Although most observations can be explained with the emission of high-energy particles that
are accelerated in the jets of AGN , the observations do not exclude that the emitting particles 
are accelerated closer to the black hole. If the magnetic field in the black hole magnetosphere has a poloidal net component on the 
order of $B_{\rm 100}\,=$ 100~G, both the spinning black hole \cite{Blandford1977} and 
the accretion disk \cite{Lovelace1976,Blandford1976} will produce strong electric fields that could 
accelerate particles to energies of $2\times 10^{19}$ $B_{\rm 100}$ eV.
High-energy protons could emit TeV photons as curvature radiation \cite{Levinson2000}, and high-energy
electrons as Inverse Compton emission \cite{2007ApJ...659.1063K}.
Such models could be vindicated by the detection of energy spectra, which are inconsistent
with originating from shock accelerated particles. An example for the latter would be 
very hard energy spectra which require high minimum Lorentz factors of accelerated particles.

The improved data from next-generation gamma-ray experiments can be compared 
with improved numerical results. The latter have recently made very substantial progress.
General Relativistic Magnetohydrodyamic codes are now able to test magnetic models of jet formation and
acceleration (see the review by \cite{2008arXiv0804.3096S}). The Relativistic-Particle-in-Cell technique 
opens up the possibility of greatly improving our understanding a wide range of issues 
including jet bulk acceleration, electromagnetic energy transport in jets, and particle
acceleration in shocks and in magnetic reconnection while
incorporating the different radiation processes \cite{Nogu:07,Nish:07,Love:05,Chan:07}.

Blazar observations would benefit from an increased sensitivity in the 
100 GeV to 10 TeV energy range to discover weaker sources and to sample the energy
spectra of strong sources on short time scales. A low energy threshold in the
10-40 GeV range would be beneficial to avoid the effect of intergalactic 
absorption that will be described further below. 
Increased sensitivity at high energies would be useful for measuring the energy spectra
of a few nearby sources like M 87, Mrk 421, and Mrk 501 at energies $\gg$10 TeV
and to constrain the effect of intergalactic absorption in the wavelength range above
10 microns. The interpretation of blazar data would benefit from dense temporal sampling
of the light curves. Such sampling could be achieved with gamma-ray experiments
located at different longitudes around the globe.

\section[Cosmic rays from star-forming galaxies]{Cosmic rays from star-forming galaxies}
\label{CR-sect}
More than 60\% of the photons detected by EGRET
during its lifetime were produced as a result of interactions
between cosmic rays (CRs) and galactic interstellar gas and dust. 
This diffuse radiation represents 
approximately $90\%$ of the MeV-GeV gamma-ray
luminosity of the Milky Way~\cite{SMR2000}. Recently H.E.S.S.
reported the detection of diffuse radiation at TeV energies
from the region of dense molecular clouds in the innermost 200\,pc around
the Galactic Center~\cite{DiffGC2006}, 
confirming the theoretical expectation
that hadronic CRs could produce VHE radiation in their interaction with
atomic or molecular targets, 
through the secondary decay of $\pi^\circ$'s. 
Only one extragalactic source of diffuse GeV radiation was
found by EGRET: the Large Magellanic Cloud, 
located at the distance of $\sim 55$\,kpc~\cite{Sreekumar1992}. 
A simple re-scaling argument suggests that a putative galaxy
with Milky-Way-like gamma-ray luminosity, located at the distance of 
1\,Mpc, would have a flux of 
approximately $2.5\times 10^{-8}$ cm$^{-2}$ s$^{-1}$ ($>100$MeV), 
well below the detection limit of EGRET and $\sim 2\times
10^{-4}$ of the Crab Nebula flux ($>1$ TeV), well below the sensitivity of
VERITAS and H.E.S.S. Thus, a next-generation gamma-ray observatory 
with sensitivity at least an order of magnitude better than VERITAS would 
allow the mapping of GeV-PeV cosmic rays in normal local 
group galaxies, such as M31, and study diffuse radiation from more distant 
extragalactic objects if their gamma-ray luminosity is enhanced by a 
factor of ten or more over that of the Milky Way.

Nearby starburst galaxies (SBG's), such as NGC253, M82, IC342, M51 exhibit
regions of strongly enhanced star formation and supernova (SN) explosions,
associated with gas clouds which are a factor of $10^{2}-10^{5}$ more dense
than the average Milky Way gas density of $\sim 1$ proton per cm$^{3}$. 
This creates nearly ideal conditions for the emission of intense, diffuse VHE
radiation, assuming that efficient hadronic CR production takes place in the
sites of the SNR's (i.e. that the galactic CR origin paradigm is valid) and in
colliding OB stellar winds~\cite{Volk1996}. In addition, leptonic 
gamma-ray production through inverse-Compton scattering of 
high density photons produced by OB associations may become effective 
in star forming regions~\cite{Pohl1994}. Multiple attempts to detect SBGs
have been undertaken by the first generation ground-based gamma-ray 
observatories. At TeV energies, M82, IC342, M81, and NGC3079 were observed 
by the Whipple 10\,m telescope~\cite{Nagai2005}, while M82 and NGC253 were 
observed by HEGRA. However, none of these objects were 
detected. A controversial detection of NGC 253 by the CANGAROO collaboration 
in 2002~\cite{Itoh2002} was ruled out by H.E.S.S. observations~\cite{HESS253}. 
The theoretical predictions of TeV radiation from starburst galaxies 
have not yet been confirmed by observations and these objects will be 
intensively studied by the current generation instruments during the next 
several years. The optimistic theoretical considerations suggest that a 
few SBG's located at distances less than $\sim 10$ Mpc may be discovered. 
Should this prediction be confirmed, a next-generation gamma-ray 
observatory with sensitivity at least an order of magnitude better than VERITAS 
will potentially discover thousands of such objects within the $\sim 100$ 
Mpc visibility range. This will enable the use of SBG's as 
laboratories for the detailed study of the SNR CR acceleration paradigm
and VHE phenomena associated with star formation, including quenching effects 
due to evacuation of the gas from star forming regions by SNR shocks 
and UV pressure from OB stars. 

If accelerated CR's are confined in the regions of high gas or photon density
long enough that the escape time due to diffusion through the magnetic field
exceeds the interaction time, then the diffuse gamma-ray flux cannot be
further enhanced by an increased density of target material, 
and instead an increased SN rate is needed. Ultra Luminous InfraRed 
Galaxies (ULIRGs), which have SN rates on the scale of a few per year 
(compared to the Milky Way  rate of $\sim 1$ per century) 
and which also have very large amounts of molecular material,
are candidates for VHE emission~\cite{Torres2004}. Although
located at distances between ten and a hundred times farther than the most
promising SBG's, the ULIRG's Arp220, IRAS17208, and NGC6240 may be within the
range of being detected by Fermi, VERITAS and 
H.E.S.S.~\cite{Torres2004a}. Next-generation gamma-ray instruments 
might be able to detect the most luminous objects of this type even if they 
are located at $\sim 1$ Gpc distances. Initial studies of the population of ULIRGs 
indicate that these objects underwent significant evolution through the
history of the Universe and that at the moderate redshift ($z<1$) the abundance 
of ULIRGs increases. Any estimate of the number of ULIRGs that may be detected is 
subject to large uncertainties due to both the unknown typical gamma-ray 
luminosity of these objects and their luminosity evolution. However, if theoretical 
predictions for Arp220 are representative for objects of this type, 
then simple extrapolation suggests $>10^{2}$ may be detectable. 

The scientific drivers to study ULIRG's are similar to those of SBGs and may include research of 
galaxy gamma-ray emissivity as a function of target gas density, 
supernova rate, confining magnetic field, etc. In addition, research of ULIRGs 
may offer a unique possibility to observe 
VHE characteristics of star formation in the context of the recent history of 
the Universe ($z<1$) since ULIRGs might be detectable to much further distances.
Other, more speculative, avenues of research may also be available. A 
growing amount of evidence suggests that AGN feedback mechanism connects episodes 
of intense starbursts in the galaxies with the accretion activity of central 
black holes. One can wonder then if a new insight into this phenomena can be 
offered by observation of VHE counterparts of these processes detected from 
dozens of ULIRGs in the range from 0.1-1 Gpc.

\section[The largest particle accelerators in the universe: radio galaxies,
galaxy clusters, and large scale structure formation shocks]{The largest particle  accelerators in the Universe: radio galaxies,
galaxy clusters, and large scale structure formation shocks}
\label{CR-clusters}
\begin{figure*}[t!h]
\begin{minipage}[!ht]{11.2cm}
\includegraphics[width=11cm]{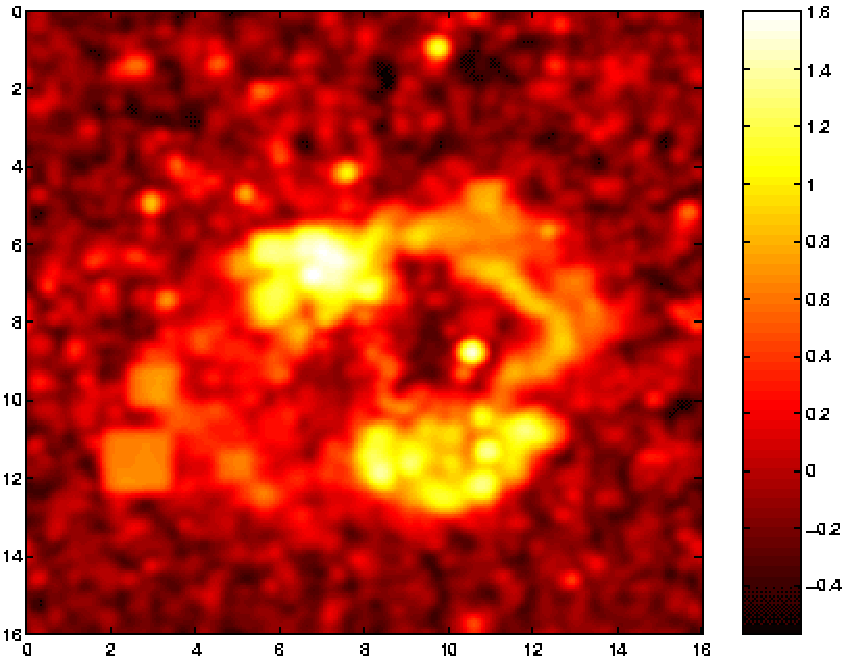}
\end{minipage}
\begin{minipage}[ht]{5cm}
\caption{\label{keshet} Results from a cosmological simulation showing how the 
$>10$ GeV gamma-ray emission from a nearby rhich galaxy cluster could look like when mapped
with a gamma-ray telescope with 0.2$^{\circ}$ angular resolution. The image covers a 
16$^{\circ}\times 16^{\circ}$ region (color scale: log($J/\bar{J}$) for an average 
$>$10 GeV flux of $\bar{J}\,=$ 8.2$\times 10^{-9}$ cm$^{-2}$ sec$^{-1}$ sr$^{-1}$) (from \cite{Keshet2002})}
\end{minipage}
\end{figure*}

%

The possibility of observing diffuse GeV and TeV radiation from even
more distant, rich galaxy clusters (GCs) has widely been discussed in the literature.
As the Universe evolves, and structure forms on increasingly larger scales, 
the gravitational energy of matter is converted into random kinetic 
energy of cosmic gas. In galaxy clusters, collisionless structure formation 
shocks, triggered by accretion of matter or mergers, are thought to be the 
main agents responsible for heating the inter-cluster medium (ICM) to 
temperatures of $\sim 10$ keV. Through these processes a fraction of 
gravitational energy is converted into the kinetic energy of 
non-thermal particles: protons and electrons. Galactic 
winds~\cite{VoelkAtoyan1999} and re-acceleration of mildly relativistic 
particles injected into the ICM by powerful cluster 
members~\cite{EnsslinBiermann1998} may accelerate 
additional particles to non-thermal energies. Cosmic ray protons can 
escape clusters diffusively only on time scales much longer than the 
Hubble time. Therefore, they accumulate over the entire formation 
history~\cite{VoelkAtoyan1999} and interact with the intercluster thermal 
plasma to produce VHE gamma radiation. Theoretical predictions for 
the detection of such systems in gamma rays by VERITAS and H.E.S.S. 
include clusters in the range from $z=0.01$ to 
$z=0.25$ (see Fig.\ \ref{keshet}) \cite{Volk1996,Keshet2002,GB2003}. 
Objects of this category were observed with Whipple \cite{Perkins2006} and H.E.S.S.
\cite{HessCluster} but were not detected. Multiple attempts to find gamma-ray 
signals from GCs in EGRET data also failed. Nevertheless, a large theoretical 
interest~\cite{BD2004,R2004,Rf2004} motivates 
further observations of the particularly promising candidates, such 
as the Coma and Perseus clusters by VERITAS and H.E.S.S.. If nearby 
representatives of the GC class are detected, a next-generation gamma-ray 
observatory with sensitivity increased by a factor of $10$   
would be able to obtain spatially resolved energy spectra from 
the close, high-mass systems, and should be able to obtain flux 
estimates and energy spectra of several dozen additional clusters. 
The detection of gamma-ray emission from galaxy clusters would make 
it possible to study acceleration mechanisms on large scales 
($>10$ kpc). It would permit measurement of the energy density of 
non-thermal particles and investigation of whether they affect the process of star 
formation in GCs, since their equation of state and cooling behavior 
differs from that of the thermal medium. If cosmic ray protons indeed 
contribute noticeably to the pressure of the ICM, measurements 
of their energy density would allow for improved estimates 
of the cluster mass based on X-ray data, and thus improve estimates 
of the universal baryon fraction. Based on population studies
of the gamma-ray fluxes from GCs, one could explore 
the correlation of gamma-ray luminosity and spectrum with 
cluster mass, temperature, and redshift. If such correlations are 
found, one could imagine using GCs as steady 
\textquotedblleft standard candles\textquotedblright \ to measure 
the diffuse infrared and visible radiation of the Universe
through pair-production attenuation of gamma rays. From a theoretical 
point of view the spectral properties of gamma-ray fluxes from GCs  
might be better understood than the intrinsic properties of blazars.  

The anticipated discovery of extragalactic sources by VERITAS and H.E.S.S. will put theoretical predictions 
discussed here on firmer ground, at least for the number of sources that the next 
generation ground-based observatory may detect. Over the next five years, 
Fermi will make major contributions to this area of studies. If the origin of 
gamma radiation in these sources is hadronic, Fermi should be able 
to detect most of the SBGs, ULIRGs, and GCs, which could potentially be 
detected by VERITAS and H.E.S.S. Under some scenarios, in which gamma rays 
are produced via leptonic mechanisms, a fraction of sources may escape Fermi 
detection (M82 might be such example), yet may still be detectable 
with VERITAS and H.E.S.S. Future theoretical effort will be required to guide
observations of these objects. In general, benefiting from the full 
sky coverage of Fermi, a program to identify the Fermi 
sources using the narrow field of view ACT observatories of the present day will
be possible, and it is likely that diffuse gamma-ray extragalactic 
sources will be discovered. Fermi will measure the galactic and extragalactic 
gamma-ray backgrounds with unprecedented accuracy and will likely resolve
the main contributing populations of sources in the energy domain 
below a few GeV. The task of determining the contribution from the diffuse 
gamma-ray sources to the extragalactic background in the range above 
a few GeV to $\sim 100$ GeV will be best accomplished by the next 
generation ground-based instrument, capable of detecting a large number of sources 
rather than a few. Most of these sources are anticipated to be weak, so they will
require deep observations. 

Large scale structure formation shocks could accelerate protons and
high-energy electrons out of the intergalactic plasma. Especially in the 
relatively strong shocks expected on the outskirts of clusters and on the 
perimeters of filaments, PeV electrons may be accelerated in substantial numbers. CMB
photons Compton scattered by electrons of those energies extend into the
TeV gamma-ray spectrum. The energy carried by the scattered photons cools
the electrons rapidly enough that their range is limited to regions close
to the accelerating shocks. However, simulations have predicted that the
flux of TeV gamma rays from these shocks can be close to detection limits
by the current generation of ground-based gamma-ray telescopes \cite{2003MNRAS.342.1009M}.
If true, this will be one of the very few ways in which these shocks can be 
identified, since very low thermal gas densities make their X-ray detection
virtually impossible. Since, despite the low gas densities involved, these 
shocks are thought to be a dominant means of heating cluster gas, their study 
is vital to testing current models of cosmic structure formation.

The origin of ultra-high-energy cosmic rays (UHECRs, $E^{>}_{\sim} 10^{16}\,\rm eV$) 
is one of the major unsolved problems in contemporary astrophysics. 
Recently, the Auger collaboration reported tentative evidence for a correlation of
the arrival directions of UHECRs with the positions of nearby Active Galactic Nuclei. 
Gamma-ray observations may be ideally suited to study the acceleration process, as the UHECRs
must produce gamma rays through various processes. The UHECRs may be accelerated far away from 
the black hole where the kpc jet is slowed down and dissipates energy.
If they are accelerated very close to the black hole at $\sim$pc distances, the high-energy particle 
beam is expected to convert into a neutron beam through photohadronic interactions 
\cite{AtoyanDermer03}. 
On a length scale $l \sim 100\, (E_n/10^{19}\,\rm eV)\,$kpc the neutron beam would
convert back into a proton beam through beta decays. 

\begin{figure}[tb]
{\epsfig{file=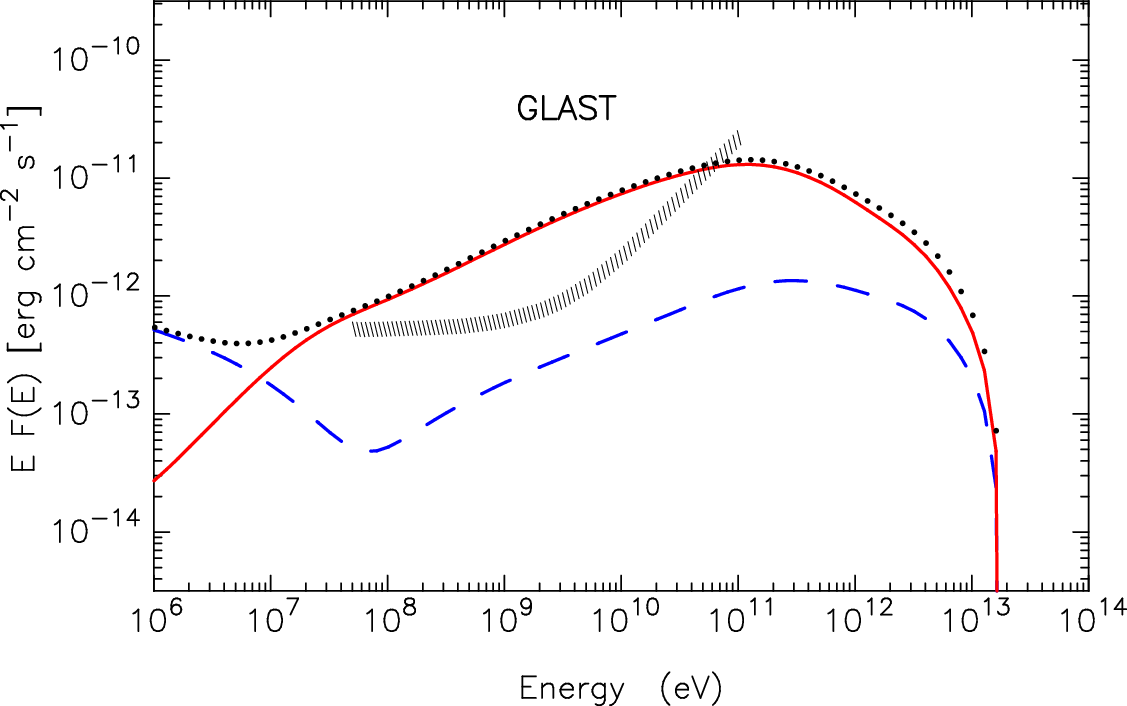,width=8cm}}
\caption{\label{cyg} Fluxes from the electromagnetic cascade initiated in  
Cyg A by UHECRs assuming the total 
injection power of secondary UHE electrons and gamma rays injected at
$\leq 1\,$Mpc distances about 
$10^{45}\, \rm erg/s$. The solid and dashed lines show the synchrotron and
Compton fluxes, respectively.}
\end{figure}

The interaction of UHECR with photons from the Cosmic Microwave Background (CMB)
creates secondary gamma rays and electrons/positrons. 
Depending on the strength of the intergalactic magnetic field ($B_{\rm IGMF}$), 
a next-generation ground-based gamma-ray experiment could detect GeV/TeV 
gamma rays from synchrotron emission of first generation electrons/positrons 
($B_{\rm IGMF}\ge 10^{-9}$~G), or inverse Compton radiation from an 
electromagnetic cascade ($B_{\rm IGMF}\le 10^{-9}$~G) \cite{2005PhRvL..95y1102G}. 
Figure \ref{cyg} shows gamma-ray fluxes expected from the electromagnetic cascade 
initiated in the CMBR and $B=3\, \mu G$ environment 
of Cyg A by injecting $10^{45}\,\rm erg/s$ of secondary electrons and/or gamma rays from 
GZK protons. For the distance to Cyg A of $\simeq 240\,\rm Mpc$
the assumed radial size of the cluster $R< \sim 1\, Mpc$ corresponds to an
extended source, or halo, of angular size $< \sim 14$ arcmin.  
Although the absorption in EBL at TeV energy is significant, the source should 
be detectable with a next-generation experiment because the source spectrum 
is very hard owing to synchrotron emission of UHE electrons. 
The detection of such emission could give information about the $\gg$TeV luminosity 
of these sources, about the intensity and spectrum of the EBL, and about the strength of the IGMF. A few aspects will be discussed further below.

A next-generation experiment might also be able to detect gamma-ray haloes with diameters 
of a few Mpc around superclusters of galaxies. Such haloes could be powered by all the sources 
in the supercluster that accelerate UHECRs. The size of the halo in these cases will be 
defined by the combination of gyroradius of the UHE electrons and their cooling path 
(synchrotron and Compton in Klein-Nishina regime). The spectral and spatial distibutions 
of such halos will contain crucial information about the EBL and 
intergalactic magnetic fields.
\section{Extragalactic radiation fields and extragalactic magnetic fields}
%
%
Very high-energy gamma-ray beams traveling over extragalactic distances
are a unique laboratory for studying properties of photons, to constrain 
theories that describe spacetime at the Planck scale and for testing radiation 
fields of cosmological origin.
The potential for probing the cosmic infrared background with TeV photons
was first pointed out by Gould and Schr\'eder \cite{Goul:67} and was revived 
by Stecker, de Jager \& Salamon \cite{Stec:92}, inspired by the detection of extragalactic TeV gamma-ray
sources in the nineties. 
High-energy gamma rays traveling cosmological  distances are attenuated 
en route to Earth by  $\gamma+\gamma \rightarrow e^+ + e^-$ interactions 
with photons from the extragalactic background light. 
While the  Universe is transparent for gamma-ray astronomy with energies below 10 GeV, 
photons with higher energy are absorbed by diffuse soft photons of wavelengths short 
enough for pair production. Photons from the EBL
in the 0.1 to 20 micron wavelength range render the Universe opaque in TeV gamma rays,  
similarly to the cosmic microwave background that constitutes a barrier for  100~TeV photons.  
The transition region from an observational window  turning opaque with 
increasing gamma-ray energy provides the opportunity for deriving observational 
constraints to the intervening radiation field.
Whereas the cosmic microwave background is accessible via direct measurements,
the cosmic infrared background (CIB) has been elusive and remains extremely 
difficult to discern by direct measurements. 
Energy spectra of extragalactic gamma-ray emitters between 10 GeV to 100 TeV allow us to 
extract information about the diffuse radiative background using spectroscopic 
measurements. Non-thermal gamma-ray emission spectra often extend over several orders of 
magnitude in energy and the high-energy absorption features expected from pair production can be 
adequately resolved with the typical energy resolution of 10\% to 20\% achievable with atmospheric
Cherenkov telescopes.

The EBL, spanning the UV to far-infrared wavelength region, consists of the cumulative 
energy releases in the Universe since the epoch of recombination  (see \cite{Haus:01} for a review). 
The EBL spectrum comprises of two distinct components. 
The first, peaking at optical to near-infrared wavelengths (0.5-2~$\mu$m), consists 
of primary redshifted stellar radiation that escaped the galactic environment either 
directly or after scattering by dust. 
In a dust-free Universe, the SED of this component can be simply determined from 
knowledge of the spectrum of the emitting sources and the cosmic history of their 
energy release. In a dusty Universe, the total EBL intensity is preserved, but the 
energy is redistributed over a broader spectrum, generating a second component 
consisting of primary stellar radiation that was absorbed and reradiated by dust 
at infrared (IR) wavelengths. This thermal emission component peaks at wavelengths 
around 100 to 140~$\mu$m. The EBL spectrum exhibits a minimum at 
mid-IR wavelengths (10 - 30 $\mu$m), reflecting the decreasing intensity of the 
stellar contribution at the Rayleigh-Jeans part of the spectrum, 
and the paucity of very hot dust that can radiate at these wavelengths. 

All energy or particle releases associated with the birth, evolution, and death 
of stars can ultimately be related to or constrained by the intensity or spectral 
energy distribution (SED) of the EBL.
The energy output from AGN represent a major non-nuclear contribution to the radiative 
energy budget of the EBL. Most of the radiative output of the AGN emerges at X-ray, UV, 
and optical wavelengths. However, a significant fraction of the AGN output can be 
absorbed by dust in the torus surrounding the accreting black hole, and reradiated at IR wavelengths. 
In addition to the radiative output from star forming galaxies and AGN, 
the EBL may also harbor the radiative imprint of a variety of ''exotic'' 
objects including Population III stars, decaying particles, and primordial
massive objects. EBL measurements can be used to constrain 
the contributions of such exotic components.\\[2ex]
%
%
Direct detection and measurements of the EBL are hindered by the fact that it has no 
distinctive spectral signature, by the presence of strong foreground emission from 
the interplanetary (zodiacal) dust cloud, and from the stars and interstellar medium of the Galaxy. 
Results obtained from TeV gamma-ray observations will complement the results 
from a number of NASA missions, i.e. Spitzer, Herschel, 
the Wide-Field Infrared Survey Explorer (WISE), 
and the James Webb Space Telescope (JWST).
In order to derive the EBL density and spectrum via gamma-ray absorption, 
ideally one would use an astrophysical standard candle of gamma rays to 
measure  the absorption component imprinted onto
the observed spectrum. In contrast, extragalactic TeV gamma-ray sources 
detected to date are highly variable AGN. Their 
gamma-ray  emission models are not unanimously agreed upon, making it 
impossible to predict the intrinsic source spectrum.
Therefore, complementary methods are required for a convincing detection of EBL attenuation.
Various approaches have been explored to constrain/measure the EBL 
intensity \cite{Stec:92,Bill:95,Vass:99,Dwek:05,Ahar:06,Mazi:07}, ranging from searching for 
cutoffs, the assumption of plausible theoretical source models, 
the possibility of using contemporaneous X-ray to TeV measurements combined 
with emission models and the concept of simultaneous constraints from direct
IR measurements/limits combined with TeV data via exclusion of unphysical gamma-ray spectra.
All of these techniques are useful; however, none has so far provided an unequivocal
result independent of assumed source spectra.

The next-generation gamma-ray experiments will allow us to use 
the flux and spectral variability of blazars \cite{Copp:99,Kren:02,Ahar:02} 
to separate variable source phenomena 
from external persistent spectral features associated with  absorption of the 
gamma-ray beam by the EBL.  
Redshift dependent studies are required to distinguish 
possible absorption by radiation fields nearby the source from extragalactic absorption. The most prominent feature of blazars is their occasional brightness (sometimes 
$>$ 10~Crab) yielding a wealth of photon statistics.  Those flares are to date
the most promising tests of the EBL density based on absorption. To constrain the EBL between 
the UV/optical all the way to the far IR a statistical sample of gamma-ray sources, and a broader 
energy coverage with properly matched sensitivity are required.

Since the cross-section for the absorption of a given gamma-ray energy is maximized
at a specific target photon wavelength (e.g., a 1 TeV gamma-ray encounters a 0.7 eV
soft photon with maximum cross-section), there is a natural division of EBL studies with 
gamma rays into three regions:  the UV to optical light, the
 near- to mid-IR and the mid- to far-IR portion of the EBL are the most effective
absorbers for $\approx$ 10 - 100 GeV, the $\approx$ 0.1 TeV to 10 TeV and the $\approx$ 
10 - 100 TeV regime, correspondingly.

In the search for evidence of EBL absorption in blazar spectra it is important 
to give consideration to the shape of the EBL spectrum showing a near IR peak, a mid IR valley 
and a far IR peak; absorption  could imprint different features onto the observed blazar spectra.
For example, a cutoff from the rapid increase of the opacity with gamma-ray energy and redshift 
is expected to be most pronounced in an energy spectral regime that corresponds to a 
rising EBL density; e.g., as is found between 0.1 - 2 micron.     This corresponds to gamma 
ray energies of 10 GeV - 100 GeV.  A survey with an instrument with sensitivity in the 10 GeV
to several 100s of GeV could measure a cutoff over a wide range of redshifts and constrain
the UV/optical IR part of the EBL.  Fermi, together with existing ground-based telescopes, is
promising in yielding first indications or maybe first conclusive results for a 
detection of the EBL absorption feature.    
However, an instrument 
with a large collection area over the given energy range by using the ground-based gamma-ray 
detection technique would allow stringent tests via spectral variability measurements. 

Similarly, a substantial rise in the opacity with gamma-ray energy is expected
in the energy regime above 20 TeV, stemming from the far IR peak.   
A corresponding cutoff should occur in the 20-50 TeV regime.   Prospective 
candidate objects are Mrk~421, Mrk~501 or 1ES1959+650, as they provide episodes of
high gamma-ray fluxes, allowing a search for a cutoff with ground-based instruments 
that have substantially enlarged collection areas in 10 - 100 TeV regime.  Sensitivity for 
detection of a cutoff in this energy regime requires IACTs with a collection area in excess 
of $\rm 1 km^{^2}$.  

Finally, a promising and important regime for ground-based telescopes to contribute
to EBL constraints lies in the  near and the mid IR (0.5 - 5 micron).   The peak in the
near IR and the slope of decline in the mid IR could lead to unique spectral imprints 
onto blazar spectra  around 1-2 TeV, assuming sufficient instrumental sensitivity.   A steep decline 
could lead to a decrease in opacity, whereas a minimal decline could result in steepening 
of the slope of the source  spectrum. If  this feature is sufficiently 
pronounced and/or the sensitivity of the instrument is sufficient, it could be a powerful method 
in unambiguously deriving the level of absorption and discerning the relative near to mid IR density.
  The location of the near IR peak and, consequently, the corresponding change in absorption, is 
expected to occur around 1.5 TeV, which requires excellent sensitivity between 100 GeV and 10 TeV.
The discovery of a signature for EBL absorption at a characteristic energy would be extremely 
valuable in establishing the level of absorption in the near to mid IR regime.  The origin of any 
signature could be tested using spectral variations in blazar spectra and discerning a 
stable component. 

A powerful tool for studying the redshift dependence of the
EBL intensity are pair haloes \cite{Ahar:94}.  For suitable IGMF strengths, such 
haloes will form around powerful emitters of $>$100 TeV gamma rays or UHECRs, 
e.g. AGN and galaxy clusters. 
If the intergalactic magnetic field (IGMF) is not too strong, the high-energy radiation 
will initiate intergalactic electromagnetic pair production and inverse Compton cascades. 
For an intergalactic magnetic field (IGMF) in the range between $10^{-12}$~G and $10^{-9}$~G
the electrons and positrons can isotropize and can result in a spherical halo glowing 
predominantly in the 100 GeV -- 1 TeV energy range. These haloes should have large
extent with radial sizes $>$ 1 Mpc. The size of a 100 GeV halo 
surrounding an extragalactic source at a distance of 1 Gpc could be 
less than 3$^{\circ}$ and be detectable with a next-generation IACT
experiment. The measurement of the angular diameter of such a halo
gives a direct estimate of the local EBL intensity at the redshift of
the pair halo. Detection of several haloes would thus allow us to
obtain unique information about the total amount of IR light produced
by the galaxy populations at different redshifts.

For a rather weak IGMF between $\sim\,10^{-16}$~G and $\sim\,10^{-24}$~G,
pair creation/inverse Compton cascades may create a GeV/TeV ''echo'' of a 
TeV GRB or AGN flare \cite{Plaga95}. The IGMF may be dominated by a primordial 
component from quantum fluctuations during the inflationary epoch of the Universe, 
or from later contributions by Population III stars, AGN, or normal galaxies. The time delay 
between the prompt and delayed emission depends on the deflection of the 
electrons by the IGMF, and afford the unique 
possibility to measure the IGMF in the above mentioned interval of 
field strengths.

\end{document}